\documentclass[superscriptaddress,showpacs,preprintnumbers,aps,twocolumn]{revtex4}

\usepackage[dvips]{graphicx}
\usepackage{amsmath}

\begin{document}

\preprint{in preparation (2005)}

\title{Counting statistics and decoherence in coupled
  quantum dots}

\author{G.~Kie{\ss}lich}
\affiliation{Institut f{\"u}r Theoretische Physik, Technische Universit{\"a}t
  Berlin, D-10623 Berlin, Germany}
\author{P.~Samuelsson}
\author{A.~Wacker}
\affiliation{Department of Physics, University of Lund, Box 118, SE-22100 Lund, Sweden}
\author{E.~Sch{\"o}ll}
\affiliation{Institut f{\"u}r Theoretische Physik, Technische Universit{\"a}t
  Berlin, D-10623 Berlin, Germany}

\begin{abstract}
We theoretically consider charge transport through two quantum dots
coupled in series. The corresponding full counting statistics for
noninteracting electrons is
investigated in the limits of sequential and coherent tunneling by means
of a master equation approach and a density matrix formalism,
respectively. We clearly demonstrate the effect of quantum coherence on the
zero-frequency cumulants of the transport process, focusing on noise and skewness.
Moreover, we establish the continuous transition from the coherent to the incoherent tunneling
limit in all cumulants of the transport process and compare this with decoherence
described by a dephasing voltage probe model. 
\end{abstract}

\maketitle


{\it Introduction}.
The analysis of current fluctuations in mesoscopic conductors provides
detailed insight
into the nature of charge transfer \cite{BLA00,NAZ03}. The complete information is available by
studying the full counting statistics (FCS), i.e. by the knowledge of all cumulants of the distribution
of the number of transferred charges \cite{LEV,NAZ03}.
As a crucial achievement, the measurement of the third-order cumulant of transport through a
single tunnel
junction was recently reported \cite{REU}.
To what extent one can extract informations from current fluctuations about
quantum coherence and decoherence is the subject of intense 
theoretical investigations: e.g. dephasing in mesoscopic cavities and
Aharonov-Bohm rings \cite{PAL04} and
decoherence in  a Mach-Zehnder interferometer \cite{FOE05}.

Quantum dots (QDs) constitute a representative system for
mesoscopic conductors. Recently, the real-time tunneling of 
single electrons could be observed  in QDs \cite{WEI}
providing an important step towards an experimental observation
of the FCS. For single dots the FCS is known to display no effects of
quantum coherence \cite{JON96,BAG03a}.
In contrast, in serially-coupled double QDs \cite{WIE03}
the superposition between states from both dots
causes prominent coherent effects. Noise properties have been
studied theoretically both in the low \cite{ELA02} and finite frequency
range  \cite{SUN99,AGU04} for these structures but
no FCS studies are available yet.
Experimentally, the low-frequency noise has been 
investigated very recently in related  double-well junctions \cite{YAU}.

In this Letter we show that detailed information about quantum
coherence in double QD systems 
can be extracted from the zero-frequency current
fluctuations. For this purpose we elaborate on the FCS
in the limits of coherent and incoherent transport through the QD
system by means of a density matrix (DM) and master equation (ME)
description. We demonstrate a smooth transition between these 
approaches by decoherence originating from coupling the QDs to a charge
detector. The results
are compared to a scattering approach, where
decoherence is introduced via phenomenological voltage probes.


{\it Model}.
The central quantity in the FCS is $P(N,t_0)$,
the distribution function of the number
$N$ of transferred charges in the time interval $t_0$. 
The associated cumulant generating function (CGF)
$F(\chi )$ is \cite{NAZ03}

\begin{eqnarray}
\exp{[-F(\chi )]}=\sum_NP(N,t_0)\exp{[iN\chi ]}
\label{eq:CGF-general}
\end{eqnarray}

\noindent
Here we consider the zero frequency limit, i.e. $t_0$ much longer
than the time for tunneling through the system.
From the CGF we can obtain the 
cumulants $C_k=-(-i\partial_\chi)^kF(\chi )\vert_{\chi=0}$
which are related to e.g. the average current $\langle I\rangle
=eC_1/t_0$ and to the zero-frequency noise $S=2e^2C_2/t_0$. The Fano
factor is defined as $C_2/C_1$. The skewness of the distribution of transferred charges
is given by the third-order cumulant $C_3$.


\begin{figure}[b]
  \begin{center}
    \includegraphics[width=0.45\textwidth]{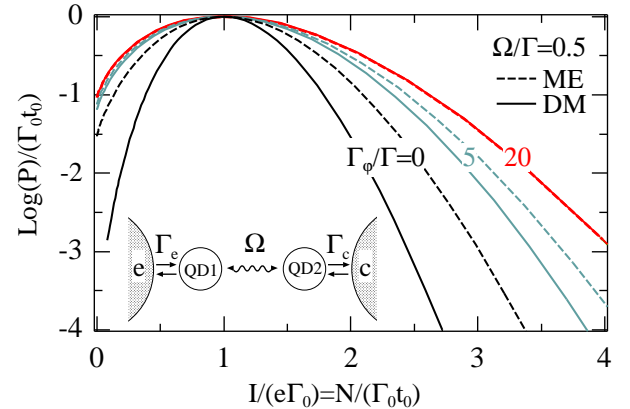}
    \caption{(Color online) Current statistics for $\Omega /\Gamma =0.5$
      and for various dephasing rates $\Gamma_\varphi
      /\Gamma =$0, 5, 20; dashed lines: master equation (ME) approach,
      solid lines: density matrix (DM) formalism; on-resonance $\Delta
      \varepsilon =0$, symmetric contact coupling: $\Gamma =\Gamma_\textrm{e}=\Gamma_\textrm{c}$.
      $\Gamma_0\equiv (2\Gamma \Omega^2)/[
      4\Omega^2+\Gamma (\Gamma +\Gamma_\varphi )]$. 
      Inset: Setup of the coupled QD system with (e)mitter and
      (c)ollector contact and mutual coupling $\Omega$.}
    \label{fig1}
  \end{center}
\end{figure}

The setup of the coupled QD system is shown as the inset of
Fig.~\ref{fig1}: 
QD1 is connected to
the emitter with a tunneling rate $\Gamma_\textrm{e}$ and QD2 to the collector contact with
rate $\Gamma_\textrm{c}$. Mutually they are coupled by the tunnel matrix element
$\Omega$. 
One level in each dot, at
energies $\varepsilon_1$ and $\varepsilon_2$ respectively, is
assumed. We consider zero temperature and work in the limit of large bias applied between the
collector and emitter, with the broadened energy levels well inside
the bias window. 
To compare DM/ME- and scattering approaches we consider noninteracting
electrons (spin degrees of freedom decouple,
we give all results for a single spin direction)
throughout this
Letter. We
note, however, that strong Coulomb blockade can be treated within the
DM/ME-approaches along the same lines.

{\it Coherent tunneling}.  
The FCS for coherent tunneling through
coupled QDs can be obtained from the approach developed by Gurvitz
and 
coworkers in a series of papers \cite{GUR96c,ELA02} (for related
work see e.g. Ref.~\cite{RAM04}). Starting from the
time dependent Schr{\"o}dinger equation one derives a modified Liouville equation,
a system of coupled first order differential equations for DM
elements $\rho_{\alpha\beta}^N(t_0)$ at a given number $N$ of
electrons transferred through the QD system at time $t_0$.  Here
$\alpha ,\beta\in \{a,b,c,d\}$, where $a,b,c$ and $d$ denote the
Fock-states $|00\rangle, |10\rangle, |01\rangle,|11\rangle$ of the
system, i.e., no electrons, one electron in the first dot, one in the
second dot, and one in each dot, respectively. The probability
distribution is then directly given by
$P(N,t_0)=\rho_{aa}^N(t_0)+\rho_{bb}^N(t_0)+\rho_{cc}^N(t_0)+\rho_{dd}^N(t_0)$. The FCS is formally
obtained by first Fourier transforming the DM elements as
$\rho_{\alpha\beta}(\chi,t_0)=\sum_{N}\rho_{\alpha\beta}^N(t_0)e^{iN\chi}$.
This
gives the Fourier transformed equation $\dot
\rho=\mathcal{L}_c(\chi)\rho$, with

\begin{eqnarray}
\mathcal{L}_c(\chi )=\left(
\begin{array}{cccccc}
-\Gamma_\textrm{e} & 0 & \Gamma_\textrm{c}e^{i\chi}  & 0 & 0 & 0\\
\Gamma_\textrm{e} & 0 & 0 & \Gamma_\textrm{c}e^{i\chi} & 0 & 2\Omega\\
0 & 0 & -2\Gamma & 0 & 0 & -2\Omega \\
0 & 0 & \Gamma_\textrm{e}  & -\Gamma_\textrm{c} & 0 & 0\\
0 & 0 & 0 & 0 & -\Gamma & -\Delta\varepsilon\\
0 & -\Omega & \Omega & 0 & \Delta\varepsilon & -\Gamma
\end{array}
\right)
\label{eq:matrix-coh}
\end{eqnarray}

\noindent
and $\rho\equiv \big(\rho_{aa},\rho_{bb},
\rho_{cc},\rho_{dd},\textrm{Re}[\rho_{bc}],\textrm{Im}[\rho_{bc}]\big)^T$,
$\Gamma\equiv (\Gamma_\textrm{e}+\Gamma_\textrm{c})/2$,
$\Delta\varepsilon\equiv\varepsilon_1 -\varepsilon_2$.

Note that the counting field $\chi$ enters the matrix elements in (\ref{eq:matrix-coh}), where an
electron jumps from QD2 into the collector contact. 
The CGF is
then obtained as the eigenvalue of $\mathcal{L}_c$ which goes to zero for $\chi=0$,
as required by probability conservation [see Eq.~(\ref{eq:CGF-general})]

\begin{eqnarray}
F_c(\chi)=\frac{t_0}{2}\left[2\Gamma-\left(p_1+2\sqrt{p_2^2+16\Gamma^2 \Omega^2
(e^{i\chi}-1)}\right)^{1/2}\right]
\label{eq:CGF-coh}
\end{eqnarray}

\noindent
with $p_1=2(\Gamma^2-4\Omega^2+\Delta\varepsilon^2)$ and
$p_2=\Gamma^2+4\Omega^2-\Delta\varepsilon^2$ 
for symmetric contact coupling 
$\Gamma_\textrm{e}=\Gamma_\textrm{c}=\Gamma$.


{\it Sequential tunneling}.
For incoherent tunneling the FCS can be obtained along similar lines 
from a ME \cite{BAG03a}
for the diagonal elements of $\rho$ as $\dot{\bar \rho}=\mathcal{L}_s\bar \rho$,
with $\bar \rho=(\rho_{aa},\rho_{bb},\rho_{cc},\rho_{dd})$. The
coefficient matrix is

\begin{eqnarray}
\mathcal{L}_s(\chi )=\left(
\begin{array}{cccc}
-\Gamma_\textrm{e} & 0 & \Gamma_\textrm{c}e^{i\chi} & 0\\
\Gamma_\textrm{e} & -Z & Z & \Gamma_\textrm{c}e^{i\chi}\\
0 & Z & -(2\Gamma +Z) & 0\\
0 & 0 & \Gamma_\textrm{e} & -\Gamma_\textrm{c}
\end{array}
\right)
\label{eq:matrix-seq}
\end{eqnarray}

\noindent
with the coupling between the single-particle states given by Fermi's golden rule:
$Z\equiv (2\vert\Omega\vert^2/\Gamma) L(\Delta\varepsilon,2\Gamma )$ with the normalized Lorentzian 
$L(x,w)\equiv [1+(2x/w)^2]^{-1}$ \cite{SPR04}. The CGF corresponds to 
the eigenvalue of the matrix (\ref{eq:matrix-seq}) which goes to
zero for $\chi =0$ and reads

\begin{eqnarray}
F_s(\chi )&=&\frac{t_0}{6}\left[(1+i\sqrt{3})q_1+(1-i\sqrt{3})q_2+6\Gamma+4Z\right],
\nonumber \\
q_{1/2}&=&\left[-u\pm\sqrt{u^2-v^3}\right]^{1/3}
\label{eq:CGF_seq}
\end{eqnarray}

\noindent
with $u=8Z^3+9Z\Gamma^2(1-3e^{i\chi})$ and $v=4Z^2+3\Gamma^2$.


{\it Results}.
The probability distributions for coherent and incoherent tunneling
obtained from the CGFs (\ref{eq:CGF-coh}) and (\ref{eq:CGF_seq}), respectively, in
a saddle-point approximation are plotted in Fig.~\ref{fig1} for 
$\Omega /\Gamma =0.5$, where the effect of coherence is most
pronounced. We see that the fluctuations are smaller in the coherent
limit, i.e. decoherence generally enhances current fluctuations. 
In the limits of small inter-dot coupling $\Omega \ll\Gamma $
one obtains a Poissonian transfer of unit 
elementary charges and for large coupling $\Omega\gg\Gamma $ the FCS of
a single QD is recovered \cite{JON96,BAG03a}. In these limits the statistics
for sequential and coherent tunneling are indistinguishable.

\begin{figure}[b]
  \begin{center}
    \includegraphics[width=0.45\textwidth]{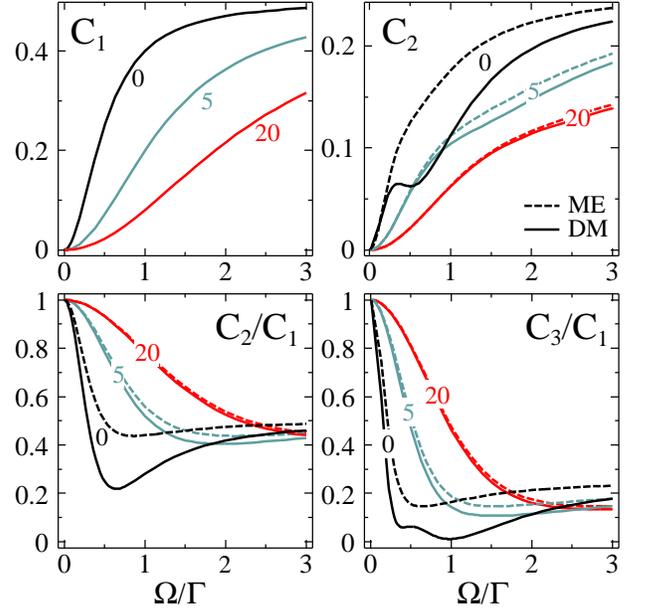}
    \caption{(Color online) Average current $C_1$, noise
      $C_2$ in units of $t_0\Gamma$, Fano factor $C_2/C_1$, normalized
      skewness $C_3/C_1$ vs. coupling $\Omega$ for
      various dephasing rates $\Gamma_\varphi /\Gamma =$0, 5, 20;
      Master equation approach (ME): dashed lines, Density matrix
      formalism (DM): solid lines. On-resonance: $\Delta\varepsilon=0$,
      symmetric contact coupling: $\Gamma =\Gamma_\textrm{e}=\Gamma_\textrm{c}$.}
    \label{fig2}
  \end{center}
\end{figure}

The CGF for coherent (\ref{eq:CGF-coh}) and sequential (\ref{eq:CGF_seq}) tunneling yield the same
expression for the average current through the coupled QD system \cite{GUR91,KOR94a,SPR04}:

\begin{eqnarray}
\langle I\rangle = e\left[\frac{1}{\Gamma_\textrm{e}}+\frac{1}{\Gamma_\textrm{c}}+\frac{1}{\Gamma_i}\right]^{-1}
L\left(\Delta\varepsilon,2\Gamma\sqrt{1+\frac{4\vert\Omega\vert^2}{\Gamma_\textrm{e}\Gamma_\textrm{c}}}\right)\nonumber\\
\label{eq:current}
\end{eqnarray}

\noindent
with $\Gamma_i\equiv 2\Omega^2/\Gamma$.
The higher order cumulants $C_k$ with
$k\geq$2 deviate for
intermediate $\Omega$ reflecting their sensitivity to quantum coherence in
the transport process. For $\Gamma=\Gamma_e=\Gamma_c$ and $\Delta\varepsilon=0$ we have
the Fano factors  \cite{SUN99,ELA02}

\begin{equation}
\frac{S_c}{2e\langle
I\rangle}=\frac{\Gamma ^4-2\Gamma ^2\Omega^2+8\Omega^4}
{(\Gamma ^2+4\Omega^2)^2}
\end{equation}

\noindent
for the coherent case and

\begin{equation}
\frac{S_s}{2e\langle
  I\rangle}=\frac{\Gamma ^4+2\Gamma ^2\Omega^2+8\Omega^4}
{(\Gamma ^2+4\Omega^2)^2}
\end{equation}

\noindent
for the sequential, incoherent case. Clearly, coherence suppresses the
noise \cite{SUN99,AGU04}. The noise and the Fano factors are
shown in Fig.~\ref{fig2} (results for
$\Gamma_{\varphi}$=0). The noise for 
coherent tunneling shows a local minimum at $2\Omega =\Gamma $. At this coupling the normalized skewness
has a local maximum as it can be seen in Fig.~\ref{fig2} and 
a close inspection reveals a FCS identical to a Poissonian transfer of
quarter elementary charges: $F(\chi )=t_0\Gamma (e^{i\chi /4}-1)$. 


{\it Decoherence - charge detector}.
In order to connect the limits of coherent and
incoherent charge transport through the QD system we consider the exponential damping of the
off-diagonal elements in the modified Liouville equation with rate $\Gamma_\varphi$: i.e. in the last two rows of the 
coefficient matrix (\ref{eq:matrix-coh}) $\Gamma$ is replaced by $\Gamma +\Gamma_\varphi$.
This apparent phenomenological treatment of decoherence can be substantiated, e.g., by the introduction
of a quantum point contact close to one of the QDs: whenever an electron enters the QD the transmission
through the quantum point contact changes. This charge detection leads to the exponential damping 
of the off-diagonals, as microscopically
derived in Ref.~\cite{GUR97}. 
Due to the finite coupling
$\Omega$, it also leads to an exponential relaxation of the diagonal
density matrix elements.
Its effect on the FCS is presented in
Fig.~\ref{fig1} and its effect on the current and noise in Fig.~\ref{fig2}. 
For comparison with the sequential tunneling cumulants
the broadening of the resonance due to the
coupling to the quantum point contact has to be considered and therefore the replacement 
$\Gamma \rightarrow \Gamma +\Gamma_\varphi$ in $Z$ of the coefficient matrix (\ref{eq:matrix-seq}) 
is carried out. Then, the currents $C_1$ in both treatments agree for any $\Gamma_\varphi$ (Fig.~\ref{fig2}).
The higher-order cumulants merge for $\Gamma_\varphi\gg\Omega$ as
shown for the noise $C_2$, the Fano factor $C_2/C_1$ and 
for the normalized skewness $C_3/C_1$ in Fig.~\ref{fig2}.

\begin{figure}[t]
  \begin{center}
    \includegraphics[width=0.45\textwidth]{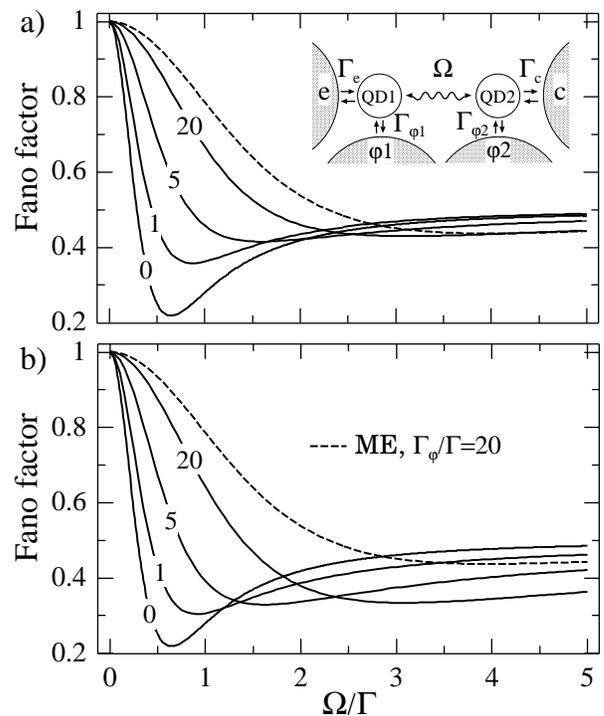}
    \caption{Fano factor vs. inter-QD coupling $\Omega$ for various dephasing rates 
        $\Gamma_\varphi$. a) elastic voltage probe, b) inelastic
        voltage probe in scattering formalism (solid curves). 
        Dashed curves: master equation (ME)  
        Fano factor for $\Gamma_\varphi /\Gamma  =20$. 
        On-resonance: $\Delta\varepsilon=0$, symmetric
        coupling: $\Gamma =\Gamma_\textrm{e}=\Gamma_\textrm{c}$.}
    \label{fig3}
  \end{center}
\end{figure}

{\it Decoherence - Voltage probe model}.
 The coherent FCS in Eq.~(\ref{eq:CGF-coh}) can also be obtained from the scattering
formula of Levitov and coworkers \cite{LEV}, $F(\chi ) =(t_0/\hbar)\int
d\varepsilon \ln[1+T(\varepsilon)(e^{i\chi}-1)]$, where
$T(\varepsilon)$ is the transmission probability through the QD
system (see e.g. \cite{ELA02}). 
This makes it interesting to compare dephasing within the DM
approach with dephasing in a scattering formalism. This is done by
introducing phenomenological voltage probes \cite{BLA00} coupled with
strength $\Gamma_{\varphi}=\Gamma_{\varphi 1}=\Gamma_{\varphi 2}$ 
to the QDs (see inset of Fig.~\ref{fig3}a). The probes absorb
and subsequently re-emit electrons, thereby randomizing their
phases. Here we focus on the current and the noise, higher cumulants
can be investigated with a modified version of the stochastic
path-integral 
technique in Ref.~\cite{PIL03}, but this is beyond the scope of the present Letter.

The scattering matrix $\mathbf{s}$ for the four terminal QD-probe system is
given by

\begin{eqnarray}
\mathbf{s}&=&1-iW^TGW, \hspace{0.2cm}G=[\varepsilon-H+iWW^T]^{-1} \\ \nonumber
H&=&\left(\begin{array}{cc} \varepsilon_1 & \Omega \\ \Omega &
\varepsilon_2 \end{array}\right), W=\left(\begin{array}{cccc}
\sqrt{\Gamma_{\varphi}} &\sqrt{\Gamma_\textrm{c}} &0 &0 \\ 0 & 0 &
\sqrt{\Gamma_{\varphi}} &\sqrt{\Gamma_\textrm{e}}\end{array}\right)
\end{eqnarray}

\noindent
The average current in lead $\alpha=e,c,\varphi_1,\varphi_2$ is given
by \cite{BLA00,BUE92}

\begin{eqnarray} 
\langle I_\alpha\rangle =\frac{e}{h}\sum_{\beta}\int d\varepsilon 
A_{\beta\beta}^\alpha (\varepsilon )f_\beta (\varepsilon )
\label{eq:scatt-current}
\end{eqnarray}

\noindent
with $A_{\beta\gamma}^\alpha (\varepsilon)=\delta_{\alpha\beta}\delta_{\alpha\gamma}-
s_{\alpha\beta}^\dagger (\varepsilon )s_{\alpha\gamma}(\varepsilon )$
and the distribution 
function $f_\alpha (\varepsilon )$ of terminal $\alpha$. The zero-frequency 
noise
between terminal $\alpha$ and $\beta$ reads \cite{BUE92,BLA00}

\begin{eqnarray}
S_{\alpha\beta}=\frac{2e^2}{h}\sum_{\gamma\delta}\int d\varepsilon
A_{\gamma\delta}^\alpha (\varepsilon )
A_{\delta\gamma}^\beta (\varepsilon)f_\gamma (\varepsilon )\big[1-f_\delta (\varepsilon )\big]
\label{eq:spectral-power}
\end{eqnarray}

\noindent
We first consider an elastic, purely dephasing voltage probe
\cite{JON}, 
where the average
current as well as the low-frequency current fluctuations into the
probe is zero at each energy. The
conservation of average current gives the average distribution
functions $f_{\varphi1/\varphi2}$. From the conservation of the
current fluctuations one obtains the fluctuating part of the
distribution functions $\delta f_{\varphi1/\varphi2}$ in terms of the
bare current fluctuations \cite{BLA00}. The total noise is then obtained as a
weighted sum of the bare current correlations in
Eq.~(\ref{eq:spectral-power}). 
It is found
that both current and noise qualitatively reproduce the DM result.
The Fano factor is plotted in Fig.~\ref{fig3}a, however, there is a
quantitative difference. Since in the DM approach, the electrons in
the dots can exchange energy with electrons at the 
quantum point contacts, the
dephasing is inelastic and a quantitative agreement with an elastic
scattering dephasing approach is not to be expected.

To account for inelastic dephasing we next consider
inelastic voltage probes which conserve only total, energy-integrated
current and fluctuations. Trying to mimic the effect of the point
contacts in the DM approach, we assume the distribution functions in
the probes to be constant, independent of energy in the entire bias
window. The average current and noise are then obtained along the same
lines as for the purely dephasing probe. We find that the average
current coincides with the DM result, the noise, however, again differs
quantitatively but not qualitatively. The Fano factor is plotted in
Fig.~\ref{fig3}b. We thus conclude that in double QD systems, dephasing in a
scattering and a DM approach yield qualitatively similar but in
general quantitatively different results.


{\it Conclusions}.
Within density matrix and master equation approaches, we have
examined the FCS for coherent and sequential charge transport through
coupled QDs. While the average currents in the two cases coincide,
all higher cumulants differ, clearly demonstrating the sensitivity of
the charge transport to quantum coherence which generally suppresses the
fluctuations. 
Coupling the QDs to a
charge detector introduces decoherence, which results in a
continuous transition from coherent to sequential tunneling.
A scattering approach, where decoherence is
introduced via phenomenological voltage probes, gives qualitatively
similar results.


{\it Acknowledgements}.
We acknowledge helpful discussions with 
S. Pilgram and M. B{\"u}ttiker. This work was supported by Deutsche Forschungsgemeinschaft
in the framework of Sfb 296 and the Swedish Research Council.


\bibliographystyle{/share/teTeX/local/revtex4/apsrev}

\end{document}